\documentclass{scrartcl}

\usepackage[backend=biber, style=ieee, citestyle=ieee]{biblatex}
\usepackage{amsmath}
\usepackage{amssymb}
\usepackage{hyperref}
\hypersetup{
	colorlinks = true,
	linkcolor=black,   
	citecolor=blue,   
	urlcolor=blue,    
	pagebackref=true,
	implicit=false,
	bookmarks=true,
	bookmarksopen=true,
	pdfdisplaydoctitle=true
}
\usepackage{cleveref}
\usepackage{url}
\usepackage{graphicx}
\usepackage{dsfont}
\usepackage{todonotes}
\usepackage{verbatim}

\def\E{\mathds E}

\usepackage{tikz}
\usetikzlibrary{fit,positioning}

\usepackage{listings}

\lstdefinelanguage{clojure}%
{morekeywords={*,*1,*2,*3,*agent*,*allow-unresolved-vars*,*assert*,*clojure-version*,*command-line-args*,%
*compile-files*,*compile-path*,*e,*err*,*file*,*flush-on-newline*,*in*,*macro-meta*,%
*math-context*,*ns*,*out*,*print-dup*,*print-length*,*print-level*,*print-meta*,*print-readably*,%
*read-eval*,*source-path*,*use-context-classloader*,*warn-on-reflection*,+,-,->,->>,..,/,:else,%
<,<=,=,==,>,>=,@,accessor,aclone,add-classpath,add-watch,agent,agent-errors,aget,alength,alias,%
all-ns,alter,alter-meta!,alter-var-root,amap,ancestors,and,apply,areduce,array-map,aset,%
aset-boolean,aset-byte,aset-char,aset-double,aset-float,aset-int,aset-long,aset-short,assert,%
assoc,assoc!,assoc-in,associative?,atom,await,await-for,await1,bases,bean,bigdec,bigint,binding,%
bit-and,bit-and-not,bit-clear,bit-flip,bit-not,bit-or,bit-set,bit-shift-left,bit-shift-right,%
bit-test,bit-xor,boolean,boolean-array,booleans,bound-fn,bound-fn*,butlast,byte,byte-array,%
bytes,cast,char,char-array,char-escape-string,char-name-string,char?,chars,chunk,chunk-append,%
chunk-buffer,chunk-cons,chunk-first,chunk-next,chunk-rest,chunked-seq?,class,class?,%
clear-agent-errors,clojure-version,coll?,comment,commute,comp,comparator,compare,compare-and-set!,%
compile,complement,concat,cond,condp,conj,conj!,cons,constantly,construct-proxy,contains?,count,%
counted?,create-ns,create-struct,cycle,dec,decimal?,declare,def,definline,defmacro,defmethod,%
defmulti,defn,defn-,defonce,defprotocol,defstruct,deftype,delay,delay?,deliver,deref,derive,%
descendants,destructure,disj,disj!,dissoc,dissoc!,distinct,distinct?,do,do-template,doall,doc,%
dorun,doseq,dosync,dotimes,doto,double,double-array,doubles,drop,drop-last,drop-while,empty,empty?,%
ensure,enumeration-seq,eval,even?,every?,false,false?,ffirst,file-seq,filter,finally,find,find-doc,%
find-ns,find-var,first,float,float-array,float?,floats,flush,fn,fn?,fnext,for,force,format,future,%
future-call,future-cancel,future-cancelled?,future-done?,future?,gen-class,gen-interface,gensym,%
get,get-in,get-method,get-proxy-class,get-thread-bindings,get-validator,hash,hash-map,hash-set,%
identical?,identity,if,if-let,if-not,ifn?,import,in-ns,inc,init-proxy,instance?,int,int-array,%
integer?,interleave,intern,interpose,into,into-array,ints,io!,isa?,iterate,iterator-seq,juxt,%
key,keys,keyword,keyword?,last,lazy-cat,lazy-seq,let,letfn,line-seq,list,list*,list?,load,load-file,%
load-reader,load-string,loaded-libs,locking,long,long-array,longs,loop,macroexpand,macroexpand-1,%
make-array,make-hierarchy,map,map?,mapcat,max,max-key,memfn,memoize,merge,merge-with,meta,%
method-sig,methods,min,min-key,mod,monitor-enter,monitor-exit,name,namespace,neg?,new,newline,%
next,nfirst,nil,nil?,nnext,not,not-any?,not-empty,not-every?,not=,ns,ns-aliases,ns-imports,%
ns-interns,ns-map,ns-name,ns-publics,ns-refers,ns-resolve,ns-unalias,ns-unmap,nth,nthnext,num,%
number?,odd?,or,parents,partial,partition,pcalls,peek,persistent!,pmap,pop,pop!,pop-thread-bindings,%
pos?,pr,pr-str,prefer-method,prefers,primitives-classnames,print,print-ctor,print-doc,print-dup,%
print-method,print-namespace-doc,print-simple,print-special-doc,print-str,printf,println,println-str,%
prn,prn-str,promise,proxy,proxy-call-with-super,proxy-mappings,proxy-name,proxy-super,%
push-thread-bindings,pvalues,quot,rand,rand-int,range,ratio?,rational?,rationalize,re-find,%
re-groups,re-matcher,re-matches,re-pattern,re-seq,read,read-line,read-string,recur,reduce,ref,%
ref-history-count,ref-max-history,ref-min-history,ref-set,refer,refer-clojure,reify,%
release-pending-sends,rem,remove,remove-method,remove-ns,remove-watch,repeat,repeatedly,%
replace,replicate,require,reset!,reset-meta!,resolve,rest,resultset-seq,reverse,reversible?,%
rseq,rsubseq,second,select-keys,send,send-off,seq,seq?,seque,sequence,sequential?,set,set!,%
set-validator!,set?,short,short-array,shorts,shutdown-agents,slurp,some,sort,sort-by,sorted-map,%
sorted-map-by,sorted-set,sorted-set-by,sorted?,special-form-anchor,special-symbol?,split-at,%
split-with,str,stream?,string?,struct,struct-map,subs,subseq,subvec,supers,swap!,symbol,symbol?,%
sync,syntax-symbol-anchor,take,take-last,take-nth,take-while,test,the-ns,throw,time,to-array,%
to-array-2d,trampoline,transient,tree-seq,true,true?,try,type,unchecked-add,unchecked-dec,%
unchecked-divide,unchecked-inc,unchecked-multiply,unchecked-negate,unchecked-remainder,%
unchecked-subtract,underive,unquote,unquote-splicing,update-in,update-proxy,use,val,vals,%
var,var-get,var-set,var?,vary-meta,vec,vector,vector?,when,when-first,when-let,when-not,%
while,with-bindings,with-bindings*,with-in-str,with-loading-context,with-local-vars,%
with-meta,with-open,with-out-str,with-precision,xml-seq,zero?,zipmap,defquery,sample
},%
   sensitive,
   alsodigit=-,%
   morecomment=[l];,%
   morestring=[b]"%
  }[keywords,comments,strings]%

\title{Techreport: Time-sensitive probabilistic inference for the edge}
\author{Christian~Weilbach, Annette~Bieniusa}

\begin{document}
\maketitle
\begin{abstract}
  In recent years the two trends of edge computing and artificial intelligence
  became both crucial for information processing infrastructures. While the
  centralized analysis of massive amounts of data seems to be at odds with
  computation on the outer edge of distributed systems, we explore the
  properties of eventually consistent systems and statistics to identify sound
  formalisms for probabilistic inference on the edge. In particular we treat
  time itself as a random variable that we incorporate into statistical models
  through probabilistic programming.
\end{abstract}

\section{Motivation}

Probabilistic inference is the process of computing a probability of an event
given prior evidence. Thus, probabilistic inference models and the data they are
conditioned on are tightly coupled. Inferring $P(Y=true|X=x)$ needs to have
evidence $x$ available. Yet, the amount of available data is not only growing,
but is also becoming more distributed due to information sharing between
different systems. For example, a lot of data is nowadays originating from
low-powered, intermittently connected remote sensors on the ``edge'' of
computational systems. Information propagates between different systems with
increasing amounts of stochastic \emph{delay}. These delays can be understood as
an intrinsic property grounded in the physically distributed nature of such
systems. While the delays accumulate additively for strongly-consistent
deterministic computational systems, we follow the intuition that statistical
inference systems could be much more robust in a coordination-free inference
setting by treating the delays as random variables. Great efforts are put into
keeping the latency low for centralized cloud services, but an increasing demand
for global probabilistic online inference requires a sound mathematical
framework for this setting. Concretely, samples or parametrized distributions of
data should be exchangeable between arbitrary systems whenever they like.

In this paper, we will investigate a formal model which takes distributed data
observation and propagation into account. The language of Bayesian statistics in
form of declarative graphical models provides an attractive starting point,
since the models are interpretable for non-experts \cite{tolpin2016design}.
Further, many state-of-the-art inference methods can be cast in a composable
Bayesian way \cite{barberBRML2012} \cite{DBLP:journals/jei/BishopN07}
\cite{Gal2016Uncertainty}. Different inference methods have become known for
efficient and effective blackbox inference, e.g. sampling methods like online
MCMC or blackbox variational inference\footnote{ SGD has also recently been
  analyzed for inference purposes \cite{mandt17}.} \cite{NIPS2013_4980}
\cite{DBLP:journals/corr/TranRB17}. However, it is not yet clear how to extend
the formalism properly to do inference in models in a distributed system.

The contributions of this paper are threefold. First we formalize the concept of
time in probabilistic inference in a way that does not require a globally
consistent physical time continuum. Then we propose a practical mechanism to
deal with time in this setting. Finally we implement this new mechanism and
demonstrate its value to inform distributed inference models.

\section{Proposal: Delay-dependent inference}

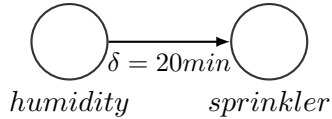
\begin{figure}
 \centering
 \begin{tikzpicture}[scale=1.8, auto, swap]
  \tikzstyle{main}=[circle, minimum size = 10mm, thick, draw =black!80, node distance = 16mm]
  \tikzstyle{connect}=[draw,-latex, thick]
  \tikzstyle{weight} = [font=\small]
  \tikzstyle{box}=[rectangle, draw=black!100]
  \node[main, fill = white!100] (beta) [label=below:$humidity$] { };
  \node[main, fill = white!10] (w) [right=of beta,label=below:$sprinkler$] { };

  \foreach \source/ \dest /\weight in {beta/w/\delta=20min}
  \path[connect] (\source) -- node[weight] {$\weight$} (\dest);
 \end{tikzpicture}
 \caption{A minimal graphical model for a sprinkler with a stochastic humidity
   sensor. To predict whether the sprinkler should turn on, the model has to
   take the delay $\delta=20min$ into account.}
 \label{fig:1}
\end{figure}

To capture the intrinsically delayed information propagation in distributed
systems, we want to extend the graphical models of Bayesian statistics with
delays. In a first step, we simply assume that the delay deltas are random
variables which are assigned to each edge connecting distributed parts of the
graph. \Cref{fig:1} shows an example of a graphical model for a sprinkler
connected to a humidity sensor. We could denote the delay explicitly with nodes
for random variables in \Cref{fig:1}, but we tie them directly to the edges
where they originate from to distinguish them from other variables. This is also
important, because the delay $\delta$ denotes the expired time since the event
happened and hence is a locally increasing clock. Whenever information is
further propagated to a remote system the transmission time is also estimated
and added. Once the event arrived the remote system will increase the delay
clock for $\delta$ in its local time again. With this clock mechanism, we define
the resulting joint distribution in factorized way by focusing on the
observation and the effect of a delay on it:

\begin{align}
 P(sprinkler|humidity, \delta) = P(sprinkler|humidity^*) \underbrace{P(humidity^*|humidity, \delta)}_{\text{delayed observation}}
\end{align}

We denote the modified observation with $humidity^*$, which can be seen as the
traditional conditional distribution.

We can pick some factorized prior $P(humidity, \delta) = P(humidity) P(\delta)$
before observations in a Bayesian way \footnote{Interarrival times in computer
  networks tend to follow a gamma distribution.}:

\begin{align}
 P(humidity) \sim Ber(0.2)                 \\
 P(\delta) \sim Gamma(k=9.0, \theta=10 min)
\end{align}

Through the prior we get a full joint distribution, that we can do Bayesian
inference on.

\subsection{Exponential decay}

A simple approach for defining $P(humidity^*|humidity, \delta)$ would be to
model it as exponential decay, where the observation fades out towards the
unobserved distribution of the observed random variable over time:

\begin{align}
 P(humidity^*=true|humidity=s, \delta=t) & = \exp(-\lambda_{\delta} t)\cdot s + (1 - \exp(-\lambda_{\delta} t)) \cdot p \label{eq:1}       \\
 P(humidity^*=false|humidity=s, \delta=t) & = 1 - P(humidity^*=true|humidity=s, \delta=t)                                           \\
 p                                               & = \underbrace{\E[humidity]}_{\text{over time}}=\frac{\sum_{s} s \Delta t_s}{\sum_s \Delta t_s}
\end{align}

This is just one possible approach to model $P(humidity^*|humidity, \delta)$.
One desired property realized in this approach is that $\delta=0$ yields the
traditional mode of observation and the observation gradually fades out towards
the prior or empirical marginal distribution of the observed variable
$humidity$. For an empirical estimate we can sum the times when the variable was
on ($s=1$) in a time interval between communicated events $\Delta t_s$ divided
by the total time the variable was observed.

The only hyperparameter is $\lambda_{\delta}$, which can be inferred if a prior
is put over it in a Bayesian setting. $\delta$ is here not a constant, but a
locally increasing clock, which will have progressed if the model is queried at
a later time locally, preferably with roughly synchronized clocks. It is
important for efficiency that this clock mechanism requires no active
computation.

\subsection{Online inference}

Once a proper model for delay-depending inference is found, the model can be run
online, where a node notifies its dependent nodes whenever its value
changes\footnote{A threshold for change might also be quantified in terms of the
  variance of the variable itself or another statistical property.}. Since the
time constant $\lambda_{\delta}$ can be inferred over time as in our experiment,
updates only have to propagate on changes relevant in its time scale. This can
provide an efficient mode of inference, where information propagation only
happens when new information is available similar to propagation
networks\cite{radul09}. The resulting system will have no driving
clock\footnote{It has no synchronization and no heart-beat mechanism.} and be
reactive\footnote{That is reacting on external observations only.}. This yields
an always-available stochastic model of computation.

\section{Related Work}

\subsection{Statistics}

A general framework to describe statistical models are graphical
forms. Graphical models allow to intuitively describe independence
assumptions about the joint distribution and the generative process behind the
data \cite{barberBRML2012}. Hierarchical graphical models allow to compose
different input sources as random variables and model complicated nested systems
of random processes including latent variables. The general formalism views the
actual inference procedure as instant, though. From a computational perspective,
the inference happens in one place at one time, appearing instantaneous to the
agent querying the model.

The graphical models assume that if random variables $X_1, ..., X_N$ are
observed, then all $N$ variables are observed at the same time and the
conditional probabilities are defined in the conventional way. The
consideration of time, either discrete or continuous, is done in the study of
stochastic processes, e.g. Markov chains. While these formalisms might provide
helpful insights, they traditionally model time as an orthogonal concept for
simplicity.

\subsection{Physics}

In special relativity, events happen in a spacetime continuum in which time is
not an orthogonal concept to spatial distribution, but modeled as a joint
Lorentzian manifold. Similarly, it is desirable that the time dimension of a
distributed network is considered as a part of the statistical manifold and not
as a separate issue, since the timing behavior in computational systems is
fairly complex and staleness can be very important for proper online
predictions.

\subsection{Distributed databases}

The crucial difference to the notion of time in distributed digital systems is
the possibility of conflicts in terms of causality. Events in replicated digital
systems can propagate in arbitrary order unless causality is modeled explicitly
\cite{DBLP:conf/sss/ShapiroPBZ11}. The general framework of eventual consistent
\footnote{i.e. converging after a finite amount of time} databases builds on
tracking this distributed discrete order of past events in lattice-like
structures and commutative algebras \cite{book:742146}
\cite{DBLP:conf/icfp/KuperN13}. For this approach of convergence the notion of
time is discrete and delays are not considered, only the order of events. While
this allows to resolve conflicts and determine a discrete digital sequential
process, the age of information can be helpful if one wants to infer the
distribution of random variables. Tracking the causal history of events might
then be avoidable, yielding a more scalable system. Furthermore a Bayesian
probabilistic framework allows to incorporate out-of-order processing in terms
of uncertainty instead of requiring explicit conflict resolution.

\subsection{Parallelized Optimization (Machine-Learning)}

Different techniques for parallelization and distribution of machine learning
algorithms have been explored, but in general they are assumed to happen in a
setting with bounded delays and most often have some form of strong centralized
coordination. These approaches do not model time as a \emph{first-class}
(explicit) concept, but rather hide it in the implicit mechanics of the
optimization algorithm, e.g. in \cite{DBLP:conf/icdcs/Teerapittayanon17}. While
the literature is increasingly rich, it does not consider a fully distributed
setting to our knowledge. It is instead focused on parallelization in
large-scale computing clusters and is seen more as an engineering problem than
as a limitation in modeling distributed inference.

\section{Experiment}

\begin{figure}
  \centering
  \includegraphics[scale=0.8]{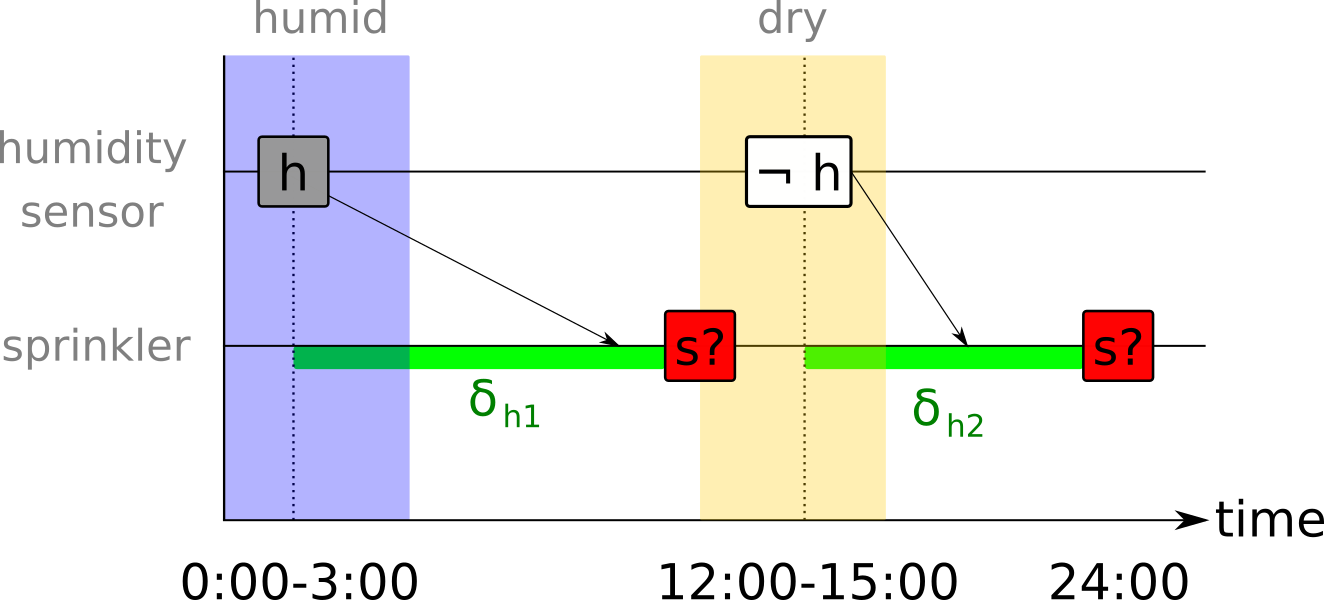}
  \caption{Schematic illustration of how the $\delta$ clocks for each
    measurement $h$ and $\neg h$ evolve. The two deltas are a summation of
    transmission time and a local clock period. The sprinkler binary variable is
    queried at noon (12:00) and at midnight (24:00). The corresponding code is
    in \Cref{code:sprinkler}.}
  \label{fig:sprinkler_experiment}
\end{figure}

We want to demonstrate with our toy example for the sprinkler with its humidity
sensor how we can use the probabilistic programming system Anglican
\cite{tolpin2016design} to model our delay mechanism in \Cref{code:sprinkler}. 

\begin{figure}
  \centering
  \begin{lstlisting}[language=clojure]
(with-primitive-procedures [exponential-decay]
  (defquery sprinkler-with-humidity-delay [data]
    (let [lambda-delta (sample (uniform-continuous 0 1.0))]
      (loop [[d & r] data]
        (if (not d) 
          lambda-delta
          (let [[s-1 s-2] d
                h-1-t (sample (uniform-continuous 0 3)) ;; 0:00 - 3:00
                h-1-delay (- 12 h-1-t)
                h-1 true ;; rained early
                h-1* (exponential-decay lambda-delta h-1-delay h-1 0.2)
                _ (observe (flip (- 1 h-1*)) s-1)

                h-2-t (sample (uniform-continuous 12 15)) ;; 12:00 - 15:00
                h-2-delay (- 24 h-2-t)
                h-2 false ;; not late
                h-2* (exponential-decay lambda-delta h-2-delay h-2 0.2)
                _ (observe (flip (- 1 h-2*)) s-2)]
            (recur r)))))))
  \end{lstlisting}
  \caption{Generative model in Anglican (\cite{tolpin2016design}). We have
    implemented our exponential decay mechanism in an external procedure. The
    query form depends on data about the desired sprinkler behaviour. The loop
    then conditions the posterior distribution over $\lambda_{\delta}$ and
    returns samples. Note that we turn on the sprinkler for s-1 and s-2 with
    inverse probability that it is humid. We have put a uniform distribution
    over $\lambda_{\delta}$ since we have no good guess what it should be except
    that values larger than 1 are unlikely. We turn the sprinkler on at 12h and
    at 24h, each time depending on whether it is humid. We also have no guess
    (yet) of how the sending time of the humidity sensor is distributed, so we
    take the uniform distribution for simplicity.}
  \label{code:sprinkler}
\end{figure}

In reality we would measure all sensor data including the delays and actual
measurements of the humidity sensor in a supervised setting and then use the
dataset for inference of $\lambda_{\delta}$. From its distribution we can immediately
infer whether the sensor information is even useful for the random variables we
observe. If we view the interaction between the random variables as stochastic
processes in time, we want to infer how long they are correlated in some way.

\begin{figure}
  \centering
  \includegraphics[scale=0.5]{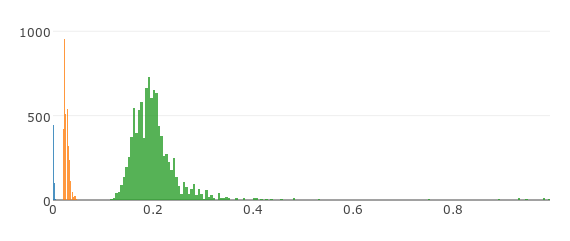}
  \caption{Using three datasets over sprinkler probabilities in blue, orange and
    green, we can infer distributions over $\lambda_{\delta}$, showing how
    quickly our observations from the humidity sensor become irrelevant. }
  \label{fig:lambda_delta_comparison}
\end{figure}

\begin{figure}
  \centering
  \includegraphics[scale=0.5]{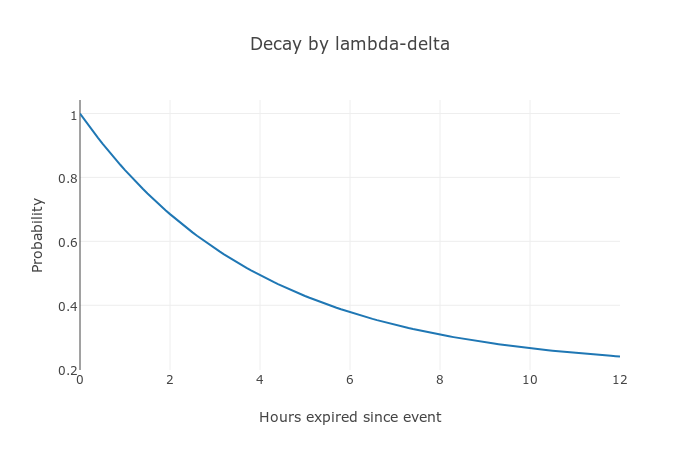}
  \caption{For $\lambda_{\delta}=0.25$ and our prior probability of humidity
    $p=0.2$, can see how our certain observation of humidity decays towards the
    prior value. At zero delay we make our observation with probability $1$,
    while after 10 hours we are close to the probability under unconditioned
    behaviour of the sprinkler. This curve matches the infered behaviour of case
    3.}
  \label{fig:exponential_decay}
\end{figure}

For demonstration purposes we have done a comparison on three synthetic datasets
of $1000$ binary sprinkler measurements as the following cases: We observe a 1.
never sprinkling at noon, always sprinkling at night with perfect correlation to
humidity 2. sprinkling at noon with probability $0.2$ and at night with
probability $0.9$ and 3. sprinkling always with the same probability $0.8$. As
can be seen by comparison in \Cref{fig:lambda_delta_comparison} there should be
no decay in information in 1. (blue) and hence $\lambda_{\delta}$ is practically
zero. For 2. (orange) the information is still valuable, but does not perfectly
describe the sprinkling behaviour. The humidity changes by factors not captured
in the sensor data. In 3. (green) the decay is so strong, that after roughly
$10$ hours it is close to the empirical average of sprinkling behaviour
(\cref{fig:exponential_decay}).

We gain several practical insights from each distribution over
$\lambda_{\delta}$. First we do not need to hardcode timing decisions in our
inference mechanism. Instead, by leveraging the standard techniques of
probabilistic inference we can immediately make an informed guess of what the
decay should be if we want to run the inference in an unsupervised setting.
Second we can evaluate how useful the sensor really is and make adjustments to
our probabilistic inference system. In case 3. we could try to run the sensor
either closer to the sprinklers query time or more often to get a better time
estimate. We can also speculate whether the humidity sensor is sending the right
signal or might be worseless. That way we could add for example additional
sensors to the system to improve its predictive power.

\section{Outlook}

The incorporation of delays into a sound mathematical framework is only a
first step towards full incorporation of time and process into graphical
models. Stochastic processes like Markov chains have been traditionally used to
describe such systems evolving in continuous time. Loops (memory) can occur in
general, complicated graphical models. Here, convergence properties of loopy
belief-propagation might be transferable. Non-stationary behavior of the delays
also needs consideration.

Once a model for distributed and decoupled statistic inference has been developed, a
universal layer of statistic inference is available. This will allow different
parties to share information with minimum effort, similar to recent open
replication systems \footnote{e.g., \url{https://ipfs.io} or
 \url{http://replikativ.io}, the latter is a project of the author.}. 

\subsection{Replication}

Furthermore, models, i.e. subgraphs, can be replicated as running, distributed
copies of the same subgraphs, and information can propagate between them. This
will require regular exchange over the inferred parameters of the underlying
statistical manifold, e.g. using some form of gossip. Maybe some lessons from
Riemannian optimization can be applied here \cite{DBLP:journals/tac/Bonnabel13}
\cite{DBLP:journals/focm/Trendafilov10}.

\subsection{Joint systems}

The composition with digital eventual consistent systems (e.g. datatypes to
collect factual data) will be challenging, as a bridge between discrete and
continuous information needs to be found to do joint computation without inducing
conflicts on the digital side.

\subsection{Differential privacy}

The resulting systems might also share sensible information between separate
parties and newly evolving methods for differential privacy will be applicable
to a distributed probabilisitic system \cite{DBLP:journals/fttcs/DworkR14}.

\subsection{Acknoweledgements}

Part of this work has been supported by the H2020 LightKone
Project\footnote{https://www.lightkone.eu/}.

\printbibliography

\end{document}